\date{}
\documentclass[showpacs]{revtex4}
\usepackage{graphicx,amsmath,amssymb,amsfonts,bbm,latexsym,color,dcolumn,epsf}

\newcommand{\dt}{\partial _t}
\newcommand{\grad}{\overrightarrow{\nabla}}

\newcommand{\rot}{\overrightarrow{\nabla}\times}

\newcommand{\A}{{\bf A}}
\newcommand{\E}{{\bf E}}
\newcommand{\B}{{\bf B}}
\newcommand{\D}{{\bf D}}
\newcommand{\K}{{\bf K}}
\newcommand{\Hc}{{\bf H}}

\newcommand{\modoescalar}{\jmath_{\ell}(\omega_{\ell k}r)Y_{\ell m}(\theta,\phi)}

\newcommand{\modoescalardt}{\jmath_ {\ell}(\frac{\jmath_{\ell p}r}{a(t)})Y_{\ell m}(\theta,\phi)}

\newcommand{\modoabrek}{\phi_{k \ell m}(\textbf{r})}
\newcommand{\s}{s^{\ell}_{p n}}

\newcommand{\gp}{g^{\ell}_{p n}}

\newcommand{\Kp}{\kappa^{2}_{\ell p}}
\newcommand{\Kn}{\kappa^{2}_{\ell n}}

\newcommand{\cteDk}{\sqrt{ \frac{2}{a_{0}^{3}}}\frac{1}{\jmath'_{\ell}(\jmath_{\ell k})}}
\newcommand{\cteDpt}{\sqrt{ \frac{2}{a^{3}(t)}}\frac{1}{\jmath'_{\ell}(\jmath_{\ell p})}}

\newcommand{\CNk}{\sqrt{ \frac{2}{a_{0}^{3}}}
\frac{1}{\jmath'_{\ell}(\kappa_{\ell
k})}\frac{1}{\sqrt{\kappa^{2}_{\ell k}-\ell(\ell+1)}}}

\newcommand{\CNpt}{\sqrt{ \frac{2}{a^{3}(t)}}
\frac{1}{\jmath'_{\ell}(\kappa_{\ell
p})}\frac{1}{\sqrt{\kappa^{2}_{\ell p}-\ell(\ell+1)}}}
\newcommand{\CNeta}{\sqrt{ \frac{2}{\textit{l}^{3}(\eta)}}
\frac{1}{\jmath'_{\ell}(\kappa_{\ell
p})}\frac{1}{\sqrt{\kappa^{2}_{\ell p}-\ell(\ell+1)}}}

\newcommand{\Wn}{\omega_{\ell n}}
\newcommand{\Wp}{\omega_{\ell p}}

\newcommand{\Wq}{\omega_{\ell q}}

\begin{document}

\title{Photon creation in a spherical oscillating cavity}

\author{Francisco D. Mazzitelli \footnote{fmazzi@df.uba.ar}}
\author{Ximena Orsi Mill\' an \footnote{xorsi@df.uba.ar}}
\affiliation{Departamento de F\'\i sica {\it Juan Jos\'e
Giambiagi}, Facultad de Ciencias Exactas y Naturales, Universidad
de Buenos Aires, Ciudad Universitaria, Pabell\' on 1, 1428 Buenos
Aires, Argentina}

%\date{today}

\begin{abstract}
We study the photon creation inside a perfectly conducting,
spherical oscillating cavity. The electromagnetic field inside the
cavity  is described by means of two scalar fields which satisfy
Dirichlet and (generalized) Neumann boundary conditions. As a
preliminary step, we analyze the dynamical Casimir effect for both
scalar fields. We then consider the full electromagnetic case. The
conservation of angular momentum of the electromagnetic field is
also discussed, showing that photons inside the cavity are created
in singlet states.
\end{abstract}


\maketitle

%%%%%%%%%%%%%%%%%%%%%%%%%%%%%%%%%%%%%%%%%%%%%%%%%%%%%%%%%%%%%%%%%%%%%%%%%%%%%%%

\section{Introduction}

The dynamical Casimir effect consists in the generation of photons
from the vacuum state of the electromagnetic field in the presence
of time-dependent boundaries or time dependent media
\cite{job,Dodonov-rev}. From the theoretical point of view, it is
widely accepted that the most favorable scenario to observe the
phenomenon involves a periodic time dependence, to enhance photon
creation by parametric resonance. Although  no concrete experiment
has yet been performed to confirm this non-stationary Casimir
effect, an experimental verification is not out of reach, and
there are several interesting proposals and ongoing experiments to
observe it \cite{prop-exp}.

Since the pioneering work by Moore \cite{moore}, the dynamical
Casimir effect for time dependent boundaries has been studied for
different fields and geometries: scalar fields in one dimensional
cavities \cite{1-dim,dodexp}, and in three dimensional rectangular
\cite{crocce,more}, and spherical \cite{setare} cavities. For the
electromagnetic field, it was analyzed in three dimensional
rectangular \cite{EM} and cylindrical cavities with time dependent
length \cite{cilindro} and radius \cite{cilindropaulo}. The
spherically symmetric situation has been considered to study
quantum radiation from a time dependent interface between two
dielectric media \cite{Eberlein}.

In this paper we consider the quantum electromagnetic field inside
a perfectly conducting, spherical cavity with a time dependent
radius. Although a spherical conducting shell may not be appealing
from an experimental point of view,  it presents many interesting
theoretical aspects. On the one hand, there is no classical
electromagnetic radiation for a pulsating sphere, not even if
charged. Thus it is interesting to check whether the quantum
effect exists or not. Moreover, the angular momentum conservation
implies that, if the effect exists, photons should be created in
singlet states. On the other hand, it is also of interest to
compare the rate of TE and TM photon creation.

The paper is organized as follows. In  Section II we describe the
classical electromagnetic field inside a spherical cavity with
time dependent radius. The TE (TM) modes are described by a scalar
field satisfying Dirichlet (generalized Neumann) boundary
conditions. For this reason, it is of interest to analyze the case
of quantum massless scalar fields satisfying both boundary
conditions, which we do in Section III. We study in detail the
resonant case in which the cavity oscillates at twice the
frequency of some field mode. In Section IV we quantize the
electromagnetic field inside the cavity, and compute the number of
TE and TM created photons. In Section V we discuss the
conservation of the angular momentum of the electromagnetic field
inside the cavity. Section VI contains our main conclusions. We
use natural units $\hbar=c=1$.

\section{Classical electromagnetic field inside a spherical cavity}

We consider a cavity bounded by a perfectly reflecting spherical
shell. We will assume that the shell is at rest for $t<0$, and
that it moves following a given trajectory $a(t)$, for $0<t<t_f$.
The trajectory is prescribed for the problem and works as a
time-dependent boundary condition for the field. Moreover, we will
assume a non relativistic motion of the shell with
$a(t)=a_0(1+\epsilon f(t))$ with $\epsilon\ll 1$ and $f(t)$ a
smooth function that vanishes for $t<0$ and $t>t_f$.

The electromagnetic field inside the cavity can be described in
terms of the four vector potential $A_{\mu}=(\varphi,{\bf A})$. In
the Coulomb gauge, $\nabla \cdot{\bf A} = 0$, the scalar potential
$\varphi$ vanishes and the vector potential satisfies the wave
equation $\Box{\bf A}=0$.

In order to obtain the vector potential $\bf A$ we consider a
function $\phi$ satisfying the scalar equation $\Box{\phi}=0$. A
solution of the wave equation for a given orientation of the axes
must be solution for any other orientation, so we can obtain $\bf A$
starting from $\phi$ by means of  ${\bf A}=\imath \bf L\phi ={\bf r}
\times \nabla\phi$. This is not the more general solution for the
electromagnetic field inside the cavity. For example, the electric
field obtained from $\bf A$ has no component along $\hat{\bf r}$ (it
is a TE mode). A linearly independent solution can be obtained,
however, by interchanging the roles of $\bf E$ and $\bf B$.
Therefore, the electromagnetic field can be written in terms of two
vector potentials ${\bf A}^{TE}$ and ${\bf A}^{TM}$. Both of the
them are obtained by the application of $\imath \bf L$ on solutions
of the scalar wave equation, $\phi^{TE}$ and $\phi^{TM}$, and
satisfy the Coulomb gauge. The complete fields are
\begin{equation}\label{Etotal}
{\bf E}={\bf E}^{TE}+{\bf E}^{TM}= -\dt{\bf A}^{TE} + \rot{\bf A}^{TM}\\
\end{equation}
\begin{equation}\label{Btotal}
{\bf B}={\bf B}^{TE}+{\bf B}^{TM}  = \rot{\bf A}^{TE} +\dt{\bf A}^{TM}\\
\end{equation}

The boundary conditions for the electromagnetic field on a moving
interface between two media are \cite{Jackson}
\begin{eqnarray}
(\D_{II}-\D_{I}).\hat{\bf n}=\sigma \nonumber
\\ (\B_{II}-\B_{I}).\hat{\bf n}=0 \nonumber
\\ \{ \hat{\bf n} \times ( \Hc_{II}- \Hc_{I})+ ( {\bf v} . \hat{\bf n} )( \D_{II}- \D_{I}) \} .
\hat{\bf t} = \K. \hat{\bf t} \nonumber
\\ \{ \hat{\bf n} \times (\E_{II}-\E_{I}) - ({\bf v}.\hat{\bf n})(\B_{II}-\B_{I})\}.\hat{\bf t}=0
\label{contornosNamias}
\end{eqnarray}
where $\hat{\bf n}$ denotes the normal to the interface going from
medium I to medium II, $\hat{\bf t}$ is any unit vector tangential
to the surface, $\sigma$ is the surface charge density and $\K$ the
surface current. These conditions can be derived using the Maxwell
equations in the laboratory frame or, alternatively, by performing a
Lorentz transformation in which a given part of the interface is
instantaneously at rest, and imposing there the static boundary
conditions \cite{mundarain}.

We assume the spherical shell to be a perfect conductor, so the
fields vanish in region II and the boundary conditions become
\begin{equation}
\B.\hat{\bf r}=0 \,\,\,\,\quad \{ \E \times \hat{\bf r} +
\dot{a}(t) \B \}.\hat{\bf t}=0 \label{contornosNamias2}
\end{equation}
For the static cavity, we have $\dot{a}(t)=0$ and the boundary
conditions are the usual ones. In terms of the scalar fields these
conditions read
\begin{equation}
\phi^{TE}|_{r=a_{0}}=0 \,\,\,\quad
[\partial_{r}(r\phi^{TM})]|_{r=a_{0}}=0 \label{cont.esc.TM}
\end{equation}
When the shell begins to move Eq. (\ref{contornosNamias2}) implies
\begin{equation}
\phi^{TE}|_{r=a(t)}=0 \,\,\,\quad \{ (\partial_{r} + \dot{a}(t)
\partial_{t}) r\phi^{TM} \}|_{r=a(t)}=0 \label{cont.esc.TM.t}
\end{equation}

From this discussion we see that the behavior of the vector
potential $\A^{TE}$ ($\A^{TM}$)is related to the problem of a
scalar field subjected to Dirichlet (generalized Neumann) boundary
conditions. The description in terms of independent TE and TM
fields is possible due to the particular geometry we are
considering. Indeed, using the above definitions and boundary
conditions it is easy to check that no mixed terms appear in
Maxwell's Hamiltonian.

The two scalar functions $\phi^{TE}$ and $\phi^{TM}$ are known as
Debye potentials \cite{Nisbet}. These functions are related with
the Hertz potentials $\Pi_{e}$ and $\Pi_{m}$ used in Ref.
\cite{cilindro} by
\begin{equation}
\Pi_{e}=  \phi^{TM} \textbf{r}  \quad\quad\quad\quad
\Pi_{m}=
\phi^{TE} \textbf{r} \label{Hertz}
\end{equation}
For the sake of simplicity, instead of dealing with the full
electromagnetic case, we will first study the problem of scalar
fields.

%***************************************************************************
\section{Quantum scalar fields inside a spherical shell }

Let us consider a massless scalar field $\phi({\bf r},t)$ inside a
spherical cavity described by the Lagrangian $\mathcal{L}$
\begin{equation}
\mathcal{L} = \frac{1}{2} \partial_{\mu}\phi \partial^{\mu}\phi
\label{lagphi}
\end{equation}
The field operator can be written in terms of creation and
annihilation operators as
\begin{equation}
\phi({\bf r},t)= \sum_{k \ell m}
 a^{in}_{k \ell m} \varphi_{k \ell m}(\textbf{r},t) + h.c.
\end{equation}
where the mode functions $\varphi_{k \ell m}({\bf r},t)$ form a
complete orthonormal set of solutions of the wave equation.

\subsection{Static cavity}

When $t<0$ we have a static spherical shell of radius $a_{0}$ and
the field modes are given by
\begin{equation}
\varphi_{k \ell m}({\bf r},t)= \modoabrek
\frac{e^{-\imath\omega_{\ell k} t}}{\sqrt{2\omega_{\ell k}}}
\label{modo.es}
\end{equation}
where
\begin{equation}
\modoabrek=\mathcal{C}_{k \ell m}\, \modoescalar
\label{autofuncion.es}
\end{equation}
are the eigenfunctions of the Laplacian with eigenvalues
$-\omega_{\ell k}^{2}$, $\,\, j_{\ell}$ are the spherical Bessel
functions and $Y_{\ell m}$ the spherical harmonics. The
normalization constants $\mathcal{C}_{k \ell m}$ depend on the
boundary conditions satisfied by the field.

We will consider two scalar fields, one satisfying Dirichlet
boundary conditions and the other one satisfying generalized Neumann
boundary conditions. It is worth to note that the scalar fields
considered here are not exactly the Debye potentials of the
electromagnetic field (the Lagrangian in Eq.(\ref{lagphi}) is not
the Maxwell Lagrangian expressed in terms of the Debye potentials).
In spite of this, we will denote them by $\phi^{TE}$ and $\phi^{TM}$
to stress the boundary conditions that they satisfy.

For the scalar field $\phi^{TE}$ the normalization constant is
given by
\begin{equation}
C_{k \ell m}= \sqrt{
\frac{2}{a_{0}^{3}}}\frac{1}{\jmath'_{\ell}(\jmath_{\ell k})}
\end{equation}
with $\jmath_{\ell k}$  the k-th zero for the spherical Bessel
function $\jmath_{\ell}(x)$. The frequencies of the modes are
$\omega_{\ell k}= \frac{\jmath_{\ell k}}{a_{0}}$

On the other hand, for the scalar field $\phi^{TM}$ we have
 \begin{equation}
C_{k\ell m}=\CNk
\end{equation}
where $\kappa_{\ell k}$ is the k-th zero of  $\{\partial_{x}[x
\jmath_{\ell}(x)]\}=0$. The TM frequencies are given by
$\omega_{\ell k}= \frac{ \kappa_{ \ell k}}{a_{0}}$.

The operators $a^{\dag in}_{k \ell m}$ and $a^{in}_{k \ell m}$
create and annihilate particles with well defined energy, total and
$z$-component of the angular momentum $\ell$ and $m$ respectively.
They correspond to the particle notion in the $"in"$ region $(t<0)$.

\subsection{Moving  cavity}

\subsubsection{Dirichlet Boundary Condition}

When the radius of the cavity depends on time, the modes of
$\phi^{TE}$ can be expanded in terms of an instantaneous basis
\begin{equation}
\varphi^{TE}_{k \ell m}(\textbf{r},t)= \sum_{p} Q^{(k)}_{p,TE}
\phi^{TE} _{p \ell m} (\textbf{r}, a(t))\label{modoDt}
\end{equation}
with
\begin{equation}
\phi^{TE} _{p \ell m} (\textbf{r}, a(t))= \cteDpt \modoescalardt
\end{equation}
Because of the spherical symmetry, in the expansion of the mode
$\varphi^{TE}_{k \ell m}(\textbf{r},t)$ it is enough to use the
functions $\phi^{TE} _{p \ell m}$ with the same values of $\ell$ and
$m$, i.e. we only mix the first quantum number. Although the
coefficients $Q^{(k)}_{p,TE}$ depend on the angular momentum $\ell$
(see Eqs.(\ref{cid}-\ref{ec.Q.TE}) below), in order to keep the
notation as simple as possible we do not write this explicitly.

The initial conditions for the coefficients $Q^{(k)}_{p,TE}(t)$ are
\begin{equation}
Q^{(k)}_{p,TE}(t=0)=\frac{\delta_{k p}}{\sqrt{2\omega_{\ell
k}}}\quad\quad \quad\quad
\dot{Q}^{(k)}_{p,TE}(t=0)=-\imath\sqrt{\frac{\omega_{\ell
k}}{2}}\delta_{k p} \label{cid}
\end{equation}
These conditions ensure that, as long as $a(t)$ and $\dot{a}(t)$
are continuous at $t=0$, each field mode and its time derivative
are also continuous functions.

The expansion in Eq.(\ref{modoDt}) for the field modes must be a
solution of the wave equation. Taking into account that at each
time the functions  $\phi^{TE} _{p \ell m} (\textbf{r}, a(t))$
form a complete and orthonormal set, the wave equation is
equivalent to the following set of coupled equations for
$Q^{(k)}_{n,TE}(t)$
\begin{eqnarray}
\ddot{Q}^{(k)}_{n,TE}(t) + [\omega_{\ell n}(t)]^{2}
Q^{(k)}_{n,TE}(t) = -2\lambda(t)\sum_{p} \dot{Q}^{(k)}_{p,TE}(t)
g^{\ell}_{p n} \nonumber \\ - \dot{\lambda}(t)\sum_{p}
Q^{(k)}_{p,TE}(t) g^{\ell}_{p n} + {\cal O}(\epsilon^2)
%\lambda^{2}(t)\sum_{p s}
%Q^{(k)}_{p,TE}(t) g^{\ell}_{p s} g^{\ell}_{n s}
\label{ec.Q.TE}
\end{eqnarray}
where $\lambda(t)=\frac{\dot{a}(t)}{a(t)}$ and
\begin{equation}
g^{\ell}_{p
n}=a(t)\int_{0}^{2\pi}\int_{0}^{\pi}\int_{0}^{a(t)}d^{3}x
\frac{\partial \phi^{TE} _{p \ell m}}{\partial a(t)} \phi^{TE
\ast} _{n \ell m} \label{G}
\end{equation}

The coefficients $g^{\ell}_{p n}$ can be computed explicitly using
that, for $k \neq k'$,
\begin{equation}
\int_{0}^{a}
r^{2}\jmath_{\ell}(kr)\jmath_{\ell}(k'r)dr=\frac{a^{2}}{k'^{2}-k^{2}}\{k\jmath_{\ell}
(k'a)\jmath'_{\ell}(ka)-k'\jmath_{\ell}(ka)\jmath'_{\ell}(k'a)\}
\label{integral}
\end{equation}
The result is
\begin{equation}
%\[
 g^{\ell}_{p n} = -g^{\ell}_{n p}= \left\lbrace
           \begin{array}{c l}
              0 & \text{if p = n}\\
              \frac{2
\jmath_{\ell n}\jmath_{\ell p}}{[\jmath_{\ell
p}]^{2}-[\jmath_{\ell n}]^{2}} & \text{if p $\neq$  n}.
           \end{array}
         \right.
         %\]
 \label{gdeD}\end{equation}

%It is interesting to note that the set of equations
%Eq.(\ref{ec.q}) is identical to the set obtained for a scalar
%field satisfying Dirichlet boundary condition in  a rectangular
%cavity with one moving wall \cite{crocce}.

\subsubsection{Neumann Boundary Condition}

We now consider a scalar field  satisfying the generalized Neumann
boundary condition on the surface of the shell. To satisfy the
boundary conditions for $t>0$ we will expand the mode functions
with respect to an instantaneous basis, as we did for the case of
Dirichlet boundary condition. However, in this case the choice for
the instantaneous basis is not so easy as before, because the
boundary condition on the moving shell given in
Eq.(\ref{cont.esc.TM.t}) involves a time derivative of the field.

The instantaneous basis can be obtained by means of a change of
variables in the $(t,r)$ plane \cite{EM}, provided that in the new
variables $( \eta, \xi )$ the boundary condition is the usual one
(i.e. no time derivative of the field)
\begin{equation}
\{ \partial_{\xi } [\xi \phi^{TM}(\eta , \xi, \theta, \varphi)]
\}|_{\xi=\textit{l} (\eta)}=0 \label{esc.Neumann.eta.xi}
\end{equation}
where $\textit{l}(\eta)$ is the value of the coordinate $\xi$ on
the moving spherical mirror.

We define the line $\eta = const$ to be a slight modification of
the line $t= const$, in such a way that it is orthogonal to the
worldline of the mirror $(t,a(t){\bf \hat r})$ at $r=a(t)$. The
coordinate $\xi$ is defined as the distance from $r=0$ to $r$ on
the line $\eta = const$. Explicitly
\begin{eqnarray}
\eta &=& t + g(r,t)\nonumber\\
\xi &=& \int^{r}_{0} dr' \sqrt{ 1 + \frac{(\partial_{r'}g
(r',t))^{2}}{ [1 + \partial_{t}g (r',t)]^{2}}}
\end{eqnarray}
where $g(r,t)=\mathcal{O}(\epsilon)$ and therefore $\xi = r +
\mathcal{O}(\epsilon^{2})$ and $l(\eta) = a(t) +
\mathcal{O}(\epsilon^{2})$.

In order to ensure the orthogonality between the line $\eta =
const$ and the world line of the mirror, we impose \cite{foot1}
\begin{equation}
g(a(t),t)=0 \quad\quad\quad
{\partial_{r}g(r,t)}|_{r=a(t)}=-\dot{a}(t) \label{condg}
\end{equation}
The function $g(r,t)$ is of course not unique. It can be expressed
as $ g(r,t)= \dot{a}(t) a(t) v(r/a(t))$, where $v(1)=0$ and
$v'(1)=-1$ (the prime denotes derivation with respect to the
argument). There are many solutions to these conditions, implying
a freedom for selecting the instantaneous basis. However, physical
quantities like the number of created particles, the energy
density inside the cavity and the angular momentum of the field
are independent of the particular choice of $g(r,t)$ \cite{EM}.

In the new coordinates, the instantaneous basis is the trivial one
\begin{equation}\label{modo.Neumann.eta.xi}
\varphi^{TM}_{k \ell m}(\eta, \xi, \theta, \varphi)= \sum_{p}
\CNeta Q^{(k)}_{p,TM}(\eta) \jmath_{\ell}(\frac{\kappa_{ \ell
p}}{\textit{l}(\eta)}\xi) Y_{\ell m}(\theta,\phi)
\end{equation}
Returning to the $(t,r)$ variables, each field mode can be
expanded as follows
\begin{equation}\label{modo.Neumann.t}
\varphi^{TM}_{k \ell m}(t,r,\theta,\varphi)= \sum_{p}
[Q^{(k)}_{p,TM}(t) + \dot{Q}^{(k)}_{p,TM}(t) g(r,t)] \phi_{p \ell
m}^{TM}(\textbf{r},a(t))+\mathcal{O}(\epsilon^{2})
\end{equation}
with
\begin{equation} \phi_{p \ell m}^{TM}(r,a(t))=  \CNpt
\jmath_{\ell}( \frac{\kappa_{\ell p}}{a(t)}r ) Y_{\ell
m}(\theta,\phi) \label{modo.abre.Neumann.t}
\end{equation}

As in the Dirichlet case, the coefficients $Q^{(k)}_{p,TM}$ depend
on the number $\ell$, but we do not write the dependence
explicitly. Assuming that $a(t)$ and $\dot{a}(t)$ are continuous
at $t=0$, and that the initial acceleration satisfies
$\ddot{a}(0)= \mathcal{O}(\epsilon^{2})$, the initial conditions
for $Q^{(k)}_{p,TM}(t)$ are the same as those for
$Q^{(k)}_{p,TE}(t)$, Eq.(\ref{cid}). The equation of motion for
$Q^{(k)}_{n,TM}(t)$ is
\begin{eqnarray}
&\ddot{Q}^{(k)}_{n,TM}(t)+[\omega_{\ell
n}(t)]^{2}Q^{(k)}_{n,TM}(t)= -2\lambda(t)\sum_{p}
\dot{Q}^{(k)}_{p,TM} \gp -\dot{\lambda}(t) \sum_{p} Q^{(k)}_{p,TM}
\gp \nonumber
\\&
-2 a^{2}(t) \dot{\lambda}(t) \sum_{p} \ddot{Q}^{(k)}_{p,TM} \s -
\sum_{p} \dot{Q}^{(k)}_{p,TM} [\s \ddot{\lambda}(t) a^{2}(t)-
\lambda(t) \eta^{\ell}_{p n}]\nonumber
\\& -\lambda(t) a^{2}(t) \sum_{p} \dt^3 Q^{(k)}_{p,TM} \s
+\mathcal{O}(\epsilon^{2})\label{ec.Q.TM}
\end{eqnarray}
where the coefficients $s^{\ell}_{p n}$ , $\eta^{\ell}_{p n}$ and
$g^{\ell}_{p n}$ are given by
\begin{equation}
s^{\ell}_{p n}=\int_{0}^{2\pi}\int_{0}^{\pi}\int_{0}^{a(t)}d^{3}x
v \phi^{TM}_{p \ell m}(\textbf{r},a(t)) \phi^{TM \ast}_{n \ell
m}(\textbf{r},a(t))
\end{equation}
\begin{eqnarray}
&\eta^{\ell}_{p
n}=\int_{0}^{2\pi}\int_{0}^{\pi}\int_{0}^{a(t)}d^{3}x
a^{2}(t)\{[\partial^{2}_{rr}v- (\omega_{\ell p})^{2} v]
\phi^{TM}_{p \ell m}(\textbf{r},a(t)) \phi^{TM \ast}_{n \ell
m}(\textbf{r},a(t))  \nonumber
\\& + \frac{2}{r}\partial_{r}v\partial_{r} [ r\phi^{TM}_{p \ell
m}(\textbf{r},a(t))] \phi^{ TM \ast}_{n \ell m}(\textbf{r},a(t))\}
\end{eqnarray}
\begin{equation}
 g^{\ell}_{p n} = \left\lbrace
           \begin{array}{c l}
              \frac{\Kp}{\Kp-l(l+1)} & \text{if p =n}\\
              \frac{\kappa_{\ell n}\kappa_{\ell p}}{(\Kn-\Kp)} \sqrt{\frac{\Kp-l(l+1)}{\Kn-l(l+1)}}
& \text{if p$\neq$  n}.
           \end{array}
         \right.
\label{gdeN}\end{equation}

\subsection{Creation of particles}

We are interested in the number of particles created inside the
cavity, so it is natural to look for harmonic oscillations of the
shell that could enhance that number by means of resonance effects
for some specific external frequencies $\Omega$. So we study the
following trajectory

\begin{equation}
a(t)= a_{0}(1 + \epsilon \sin (\Omega t))
\end{equation}

Let us first consider the Dirichlet scalar field $\phi^{TE}$. When
$t>t_{f}$, ($"out"$ region), the radius returns to its initial value
$a_{0}$,  the right hand side in Eq.(\ref{ec.Q.TE}) vanishes and the
solution is
\begin{equation}
Q^{(k)}_{n,TE}(t>t_{f})= A^{(k) }_{\ell n,TE} e^{\imath
\omega_{\ell n} t} + B^{(k) }_{\ell n,TE} e^{- \imath \omega_{\ell
n} t} \label{sol.Q.TE.final}
\end{equation}
where $A^{(k) }_{\ell n,TE}$ and $B^{(k) }_{\ell n,TE}$ are
constant coefficients to be determined by the continuity
conditions at $t=t_{f}$. In these coefficients we write the
dependence on $\ell$ explicitly.

For $t>t_{f}$ we can define a new set of operators $a^{out}_{k
\ell m}$ and $a^{\dag out}_{k \ell m}$, associated with the
particle notion in the $"out"$ region . The $"in"$ and $"out"$
operators are connected by means of the Bogoliubov
transformation
\begin{equation} a^{out} _{n \ell m}=  \sum_{k} [
B^{(k) }_{\ell n,TE} a^{in} _{k \ell m} + (-1)^{m} A^{(k) \ast
}_{\ell n,TE} a^{in \dag} _{k \ell -m} ]\sqrt{2\omega_{\ell
n}}\label{Bogoliubov}
\end{equation}
The number of $"out"$ particles in the mode $(n,\ell ,m)$ is given
by
\begin{equation}
<\mathcal{N}_{n \ell m}>=<0_{in}|a^{out \dag} _{n \ell m}a^{out}
_{n \ell m}|0_{in}> = 2 \omega_{\ell n} \sum_{k} |A^{(k) }_{\ell
n,TE}|^{2} \label{particulascreadas.M}
\end{equation}

To obtain the coefficients $A^{(k) }_{\ell n,TE}$ and $B^{(k)
}_{\ell n,TE}$ we must solve Eq.(\ref{ec.Q.TE}) for the
coefficients $Q^{(k)}_{n,TE}(t)$. They are of the same form as
those that describe the modes of a scalar field satisfying
Dirichlet boundary conditions in a three-dimensional rectangular
cavity \cite{crocce}, and can be solved using Multiple Scale
Analysis (MSA), see for example, Refs.\cite{Bender,crocce}. We
will review here the main results and include the details in the
Appendix.

The solution for the coefficients $A^{(k) }_{\ell n,TE}$ and $B^{(k)
}_{\ell n,TE}$ depends on the relation between the external
frequency $\Omega$ and the natural frequencies of the field in the
cavity. There is parametric resonance when the external frequency
$\Omega$ equals the sum of two eigenfrequencies of the cavity with
the same angular momentum $\Omega=\Wn +\Wp$. For simplicity, in what
follows we will analyze the particular case $\Omega = 2\Wn$. Using
MSA, one can show that the modes $(\ell,n)$ and $(\ell,q)$ are
coupled if any of the following conditions is satisfied
\begin{equation}
\Omega =\Wn -\Wq \quad\quad \Omega =-\Wn +\Wq \label{acopl.n}
\end{equation}

For $\Phi^{TE}$ the frequencies are given by the zeros of the
spherical Bessel functions ($\omega_{\ell n}= \frac{\jmath_{\ell
n}}{a_{0}}$) and so, the spectrum is qualitatively different
depending on the value of the number $\ell$. For $\ell=0$ the
spectrum is equidistant. In the particular case $\Omega= 2
\omega_{0n}$, the set of equations for the coefficients $A^{(k)
}_{0n,TE}$ and $B^{(k) }_{0n,TE}$ corresponds to that of a
one-dimensional cavity excited with twice the lowest eigenfrequency.
The equations are coupled, the number of particles grows linearly in
time, and the energy increases exponentially \cite{dodexp}.

On the other hand, for $\ell\neq 0$ the spectrum is not equidistant
\cite{foot2}, and one can check that there is no coupling between
modes. Eq.(\ref{ec.Q.TE}) reduces to the Mathieu equation for the
modes with frequency $\omega_{\ell n}$. The Bogoliubov coefficients
$A^{(k) }_{\ell n,TE}$ and $B^{(k) }_{\ell n,TE}$ and the number of
created particles grows exponentially.

For the case of the scalar field $\phi^{TM}$ the situation is
similar. The set of Eqs.(\ref{ec.Q.TM}) has the same form as the
set that describes the modes of a scalar field in a
three-dimensional rectangular cavity satisfying generalized
Neumann boundary conditions \cite{EM}, and can be solved again
using MSA. As before, the spectrum is equidistant for $\ell=0$ and
non-equidistant for $ \ell \neq 0$.

In the resonant case $\Omega=2\omega_{L N}$ (L $\neq$ 0), for both
TE and TM modes the number of created particles is given by (see
Appendix)
\begin{equation}
<\mathcal{N}_{N L m}>=<0_{in}|a^{out \dag} _{N L m}a^{out} _{N L
m}|0_{in}>= \sinh^{2}(\gamma \epsilon t_{f})
\label{particulascreadas.m}
\end{equation}
and
\begin{equation}
<\mathcal{N}_{N L}>=\sum_{m}<0_{in}|a^{out \dag} _{N L m}a^{out}
_{N L m}|0_{in}>= (2L+1)\sinh^{2}(\gamma \epsilon
t_{f})\label{particulascreadas}
\end{equation}
The constant $\gamma$ determines the rate of growth. For TE modes we
have
\begin{equation}
\gamma^{TE} =\frac{\jmath_{L N}}{2 a_{0}}\label{tasa.D}
\end{equation}
while for TM modes
\begin{equation}
\gamma^{TM}= \frac{\kappa_{L  N}}{2 a_{0}}
\frac{1+\frac{L(L+1)}{\kappa_{L N}^{2}}}{1 -
\frac{L(L+1)}{\kappa_{L N}^{2}}}\label{tasa.N}
\end{equation}

The case $\ell =0$ is qualitatively different and, as stated
above, equivalent to the one dimensional dynamical Casimir effect.
However, as we will see in the next section, these modes are
absent for the electromagnetic field.

%%%%%%%%%%%%%%%%%%%%%%%%%%%%%%%%%%%%%%%%%%%%%%%%%%%%%%%%%%%%%%%%%%%%%%%%%

\section{The electromagnetic field}

In Section II we showed that the electromagnetic field inside the
spherical cavity can be described in terms of two vector
potentials. They can be expanded in terms of creation and
annihilation operators as
\begin{equation}
\A^{TE}(\textbf{r},t)= \sum_{k \ell m}
 a^{in}_{k \ell m} \A^{TE}_{k \ell m}(\textbf{r},t) + h.c.
\label{ATE}
\end{equation}
\begin{equation}
\A^{TM}(\textbf{r},t)= \sum_{k \ell m}
 a^{in}_{k \ell m} \A^{TM}_{k \ell m}(\textbf{r},t) + h.c.
\label{ATM}
\end{equation}
The modes $\A^{TE}_{k \ell m}(\textbf{r},t)$ and $\A^{TM}_{k \ell
m}(\textbf{r},t)$ can be obtained from the modes of two scalars
fields through the application  of the operator $ \imath
\textbf{L} =\textbf{r} \times \grad$. This operator acts only on
the angular part of the scalar modes, so we introduce the
vectorial spherical harmonics $\textbf{X}_{\ell m}(\theta,\phi)$
\begin{equation}
\textbf{X}_{\ell m}(\theta,\phi) =
\frac{\textbf{L}}{\sqrt{\ell(\ell+1)}}Y_{\ell
m}(\theta,\phi)\label{arm.vec.}
\end{equation}
(the  additional factor $\sqrt{\ell(\ell+1)}$ is needed for
normalization). Therefore the modes for the vector potentials are
given by
\begin{eqnarray}
&  \A_{k \ell m}^{TE}(\textbf{r},t<0)= \cteDk \jmath_
{\ell}(\frac{\jmath_{\ell k}r}{a_{0}}) \textbf{X}_{\ell
m}(\theta,\phi) \nonumber
\\&
\A_{k \ell m}^{TM}(\textbf{r},t<0)= \CNk \jmath_
{\ell}(\frac{\kappa_{\ell k}r}{a_{0}}) \textbf{X}_{\ell
m}(\theta,\phi)
\end{eqnarray}
for the static cavity and
\begin{eqnarray}
& \A^{TE}_{k \ell m}(\textbf{r},t>0)= \sum_{p} {\sqrt{
\frac{2}{a^{3}(t)}}\frac{1}{\jmath'_{\ell}(\jmath_{\ell p})}}
Q^{(k)}_{p,TE}(t) \jmath_ {\ell}(\frac{\jmath_{\ell p}r}{a(t)})
\textbf{X}_{\ell m}(\theta,\phi)\nonumber
\\&
\A_{k \ell m}^{TM}(\textbf{r},t>0)= \sum_{p}\CNpt
[Q^{(k)}_{p,TM}(t) + g(\textbf{r},t) \dot{Q}^{(k)}_{p,TM}(t) ]
\jmath_ {\ell}(\frac{\kappa_{\ell p}r}{a(t)}) \textbf{X}_{\ell
m}(\theta,\phi)
\end{eqnarray}
for the moving cavity. It is worth to remark that, as
$\textbf{L}Y_{00}=0$, there is no monopolar term in the expansions
for the electromagnetic field.

The dynamical evolution of the TE (TM) modes is that of the modes of
$\phi^{TE} (\phi^{TM})$ with $\ell \neq 0$. As a consequence, the
number of created photons in each mode equals the number of created
particles of the corresponding scalar field. If we consider again
the parametric resonant case $\Omega=2 \omega_{L N}$, the number of
photons grows exponentially and is given by the
Eqs.(\ref{particulascreadas.m},\ref{tasa.D}) in the case of TE modes
and by Eqs. (\ref{particulascreadas.m},\ref{tasa.N}) in the case of
TM modes. It is interesting to note that it is in general not
possible to excite at the same time both a TE and a TM mode (for
that to be possible one should have $\omega^{TE}_{\ell
n}+\omega^{TE}_{\ell n'} =\omega^{TM}_{\ell' k}+\omega^{TM}_{\ell'
k'}$, which is not satisfied).

%%%%%%%%%%%%%%%%%%%%%%%%%%%%%%%%%%%%%%%%%%%%%%%%%%%%%%%%%%%%%%%%%%%%%%%%

\section{ANGULAR MOMENTUM}

In this section we discuss the conservation of the angular
momentum of the electromagnetic field inside the spherical cavity,
showing that photons are created in singlet states. As we start in
the vacuum state of the electromagnetic field, the average value
of the angular momentum is initially zero. The oscillations of the
radius of the cavity does not break the symmetry under rotations,
so the angular momentum must vanish at all times.

The angular momentum of the electromagnetic field is
\begin{equation}
\textbf{\L} = \int_{sphere} \textbf{r} \times (\E \times \B)
d^{3}x =\textbf{\L}^{TE} + \textbf{\L}^{TM}
 \label{Lem}
\end{equation}
As in resonant situations it is possible to produce either TE or
TM photons, in what follows we will consider only one of the two
polarizations, without specifying which one. Both
$\textbf{\L}^{TE}$ and $\textbf{\L}^{TM}$ are of the form
\begin{eqnarray}
&\textbf{\L}= (\text{\L}_{x},\text{\L}_{y},\text{\L}_{z})= \sum_{k
\ell m}(\frac{1}{2} [C^{+}_{\ell  m-1} \emph{a}^{in \dag}_{k \ell
m}{a}^{in}_{k \ell m-1} +C^{-}_{\ell  m}\emph{a}^{in \dag}_{k \ell
m-1}{a}^{in}_{k \ell m}],\nonumber
\\&\frac{-\imath}{2}[
C^{+}_{\ell m} \emph{a}^{in \dag}_{k \ell m+1}{a}^{in}_{k \ell m}
-C^{-}_{\ell m}\emph{a}^{in \dag}_{k \ell m-1}{a}^{in}_{k \ell m}]
,m \emph{a}^{in \dag}_{k \ell m}{a}^{in}_{k \ell m})
 \label{L}
\end{eqnarray}
with
\begin{equation}
\\C^{+}_{\ell m}=\sqrt{(\ell-m)(\ell+m+1)}\quad\quad\quad\quad\quad\quad
\\C^{-}_{\ell m}=\sqrt{(\ell+m)(\ell-m+1)}
\end{equation}
The expression of $\textbf{\L}$ in terms of the $"in"$ creation
and annihilation operators is valid for all times. Furthermore, as
we work in Heisenberg picture, the state of the electromagnetic
field is always the $"in"$ vacuum, so we have
\begin{equation}
<\textbf{\L}>=<0_{in}|\textbf{\L}|0_{in}>=0 \quad\quad \forall t
\end{equation}
\medskip

However, in the $"out"$ region, when  the radius of the cavity
returns to its original value and remains at rest, we could expand
the field in terms of the new set of the $"out"$ creation and
annihilation operators. Therefore the angular momentum of the field
can be written as in Eq.(\ref{L}) but changing the $"in"$ operators
by the $"out"$ ones. As the number of $"out"$ particles is different
from zero, photons must be created forming singlet states, in such a
way that the angular momentum of the field remains null.

The state of the field $|0_{in}>$ can be written as a linear
combination of $"out"$ states
\begin{equation}
|0_{in}>= \alpha |0_{out}> +  \sum_{n \ell m} \alpha_{n \ell m}
\textit{a}^{out \dag}_{n \ell m}|0_{out}> + \sum_{n \ell
m}\sum_{n' \ell' m'} \alpha_{n \ell m, n' \ell' m'}
\textit{a}^{out \dag}_{n \ell m}\textit{a}^{out \dag}_{n' \ell'
m'}|0_{out}>+..... \label{com.lineal}
\end{equation}

In the parametric case $\Omega=2\omega_{L N}$, the relation
between the $"in"$ and $"out"$ operators is, for $\ell = L$ and
$n=N$,
\begin{equation}
a^{out} _{N L m}=  [ B^{(N) }_{L N} a^{in} _{N L m} +
(-1)^{m}(A^{(N) }_{L N})^{\ast} a^{in \dag} _{N L -m}
]\sqrt{2\omega_{L N}} \label{a.out-a.in2}
\end{equation}
\begin{equation}
a^{out \dag} _{N L m}=  [ (B^{(N) }_{L N})^{\ast} a^{in \dag} _{N
L m} + (-1)^{m} A^{(N) }_{L N} a^{in } _{N L -m} ]\sqrt{2\omega_{L
N}}\label{a.outdaga-a.in2}
\end{equation}
The $"in"$ and $"out"$ operators coincide $a^{out} _{n \ell m}=
a^{in} _{n \ell m}$ when $\ell\neq L$ or $n\neq N$. Therefore, we
rewrite Eq.(\ref{com.lineal}) in the following way
\begin{equation}
|0_{in}>= \alpha |0_{out}> +  \sum_{m} \alpha_{m} |1_{out}>_{m} +
\sum_{m}\sum_{m'} \alpha_{m,m'} |1_{out}>_{m}|1_{out}>_{m'}+.....
\label{com.lineal.m2}
\end{equation}
where we omitted the subindexes  $N$ and $L$.

In order to find the $\alpha$ coefficients we apply a destruction
$"in"$ operator on both sides of Eq. (\ref{com.lineal.m2}). The
left hand side gives zero. On the right hand side we write the
$"in"$ operator in terms of the $"out"$ operators, by inverting
Eqs.(\ref{a.out-a.in2})-(\ref{a.outdaga-a.in2}). Doing that we
obtain a linear combination of orthogonal states which equals
zero, so the coefficient of each state must vanish. In this way we
get the equations that determine the $\alpha$ coefficients. In the
particular case of $\Omega=2\omega_{L N}$ with $L$=1, the state of
the field $|0_{in}>$ can be written as
\begin{eqnarray}
& |0_{in}>= A \{ |0_{out}>  - \mathcal{C} [
|1_{out}>_{1}|1_{out}>_{-1}- \frac{1}{\sqrt{2}} |2_{out}>_{0}] +
\nonumber
\\&  \mathcal{C}^{2} [|2_{out}>_{1}|2_{out}>_{-1}-
 \frac{1}{\sqrt{2}}|1_{out}>_{1}|2_{out}>_{0}|1_{out}>_{-1} +
\frac{\sqrt{3}}{2\sqrt{2}}|4_{out}>_{0}] -\nonumber
\\& \mathcal{C}^{3}
[|3_{out}>_{1}|3_{out}>_{-1} -
 \frac{1}{\sqrt{2}}|2_{out}>_{1}|2_{out}>_{0}|2_{out}>_{-1} +
\nonumber \\& \frac{\sqrt{3}}{2\sqrt{2}}
|1_{out}>_{1}|4_{out}>_{0}|1_{out}>_{-1} -
\frac{\sqrt{5}}{4}|6_{out}>_{0}] +...\} \label{com.lineal.def}
\end{eqnarray}
where $A$ is the normalization constant of the state and
\begin{equation}
\mathcal{C}=-\tanh(\gamma \epsilon t_{f})\,\, . \label{coefC}
\end{equation}
The constant $\gamma$ is given by Eq.(\ref{tasa.D}) for TE photons
and by Eq.(\ref{tasa.N}) for TM photons, both with $L=1$. Each
state between brackets in Eq. (\ref{com.lineal.def}) is an
eigenstate of $\textbf{\L}^{2}$ and $\text{\L}_{z}$ with
eigenvalue equal to zero, as expected.

%%%%%%%%%%%%%%%%%%%%%%%%%%%%%%%%%%%%%%%%%%%%%%%%%%%%%%%%%%%%%%%%%%%%%%%%

\section{CONCLUSIONS}

In this paper we have computed the resonant photon creation inside
a spherical  oscillating cavity taking into account the vector
nature of the electromagnetic field. We described the TE and TM
modes of the electromagnetic field by  massless scalar fields
satisfying Dirichlet and generalized Neumann boundary conditions.
We first studied the creation of particles for these scalar
fields, and then showed that the number of created photons in each
mode (TE or TM) equals the number of created particles of the
corresponding scalar field. We used MSA to take into account
resonant effects at long times, and found an exponential growth in
the number of created photons. Previous works studied the case of
a Dirichlet scalar field only in the short time limit
$\epsilon\Omega t << 1$ \cite{setare}. Our results are consistent
with those in this limit.

The spectrum of the scalar fields is equidistant for $\ell=0$ and
non equidistant for $\ell \neq 0$. When the external frequency is
chosen to produce parametric resonance in modes with $\ell=0$, an
infinite number of modes are excited. The problem becomes
equivalent to the one dimensional dynamical Casimir effect. The
modes with $\ell\neq 0$ are non equidistant, and there is no mode
coupling. In the parametric resonance case $\Omega = 2 \Wn$, the
number of motion-induced particles grows exponentially in the
particular modes corresponding to these values of $n$ and $\ell$,
for all possible values of $m$.

The eigenfrequencies of the TE and TM modes of the electromagnetic
field are the same as the eigenfrequencies of $\phi^{TE}$ and
$\phi^{TM}$ (the only difference is that in the electromagnetic case
the $\ell =0$ modes are absent). The growth rate is different for TE
and TM modes. In cubic cavities, where the frequencies of the TE and
TM modes are equal, they satisfy \cite{EM}
\begin{equation}
\gamma^{TE} < \frac{\omega}{2} \quad\quad \gamma^{TM} >
\frac{\omega}{2}
\end{equation}
For spherical cavities, the TE and TM modes have different
eigenfrequencies. The growth rates satisfy
\begin{equation}
\gamma^{TE} = \frac{\omega^{TE}}{2} \quad\quad \gamma^{TM}
> \frac{\omega^{TM}}{2}
\end{equation}
Therefore the ratio between the rate of growth and the
eigenfrequency is, as in cubic cavities, larger for TM modes than
for TE modes. In other words, the functional dependence of the rate of
growth with the eigenfrequency is different for
TE and TM modes.

Finally, we considered the conservation of the  angular momentum of
the electromagnetic field in the oscillating cavity. As the initial
state of the field is the vacuum, and the motion of the shell does
not break the symmetry of rotation, the  angular momentum of the
field must be zero for all time. Therefore, particles must be
created in singlet states. For the particular case $\Omega= 2
\omega_{1N}$ we wrote explicitly the $"in"$ vacuum state as a linear
combination of $"out"$ singlet states.

In classical electromagnetism there is no electromagnetic
radiation with spherical symmetry. Therefore it was not completely
obvious that the dynamical Casimir effect would occur in a
spherical oscillating cavity. A mechanical analogue of the one
dimensional dynamical Casimir effect is useful to illustrate this
point \cite{geometrias}. At the classical level it is not possible
to amplify transversal oscillations on a string by changing its
length, unless there is an initial classical wave on it
\cite{Havelock,geometrias}. At the quantum level, the initial
conditions of the modes are non trivial due to Heisenberg
uncertainty principle, and therefore it is possible to excite the
system even starting from the ground state. Analogously, in the
spherically symmetric case, the dynamical Casimir effect is non
trivial because all modes with $\ell\neq 0$ have non vanishing
quantum fluctuations.

%%%%%%%%%%%%%%%%%%%%%%%%%%%%%%%%%%%%%%%%%%%%%%%%%%%%%%%%%%%%%%%%%%%%%%%%%%%%

\section{ACKNOWLEDGEMENTS}

This  work  was supported by Universidad de Buenos Aires, Conicet,
and Agencia Nacional de Promoci\'on Cient\'\i fica y
Tecnol\'ogica, Argentina.

%%%%%%%%%%%%%%%%%%%%%%%%%%%%%%%%%%%%%%%%%%%%%%%%%%%%%%%%%%%%%%%%%%%%%%%%

\section{APPENDIX}

In this Appendix  we solve  Eqs.(\ref{ec.Q.TE}) for an oscillating
motion of the shell given by $a(t)=a_{0}(1 + \epsilon \sin(\Omega
t ))$.

For small amplitudes ($\epsilon \ll 1$), Eqs. (\ref{ec.Q.TE}) take
the form
\begin{eqnarray}
&\ddot{Q}^{(k)}_{n,TE}(t) + [\omega_{\ell n}]^{2}
Q^{(k)}_{n,TE}(t)=2\epsilon \sin( \Omega t)[\omega_{\ell n}]^{2}
Q^{(k)}_{n,TE}(t) \nonumber
\\ &+ 2\epsilon\Omega \cos(\Omega t) \sum_{p} \dot{Q}^{(k)}_{p, TE}(t)
g^{\ell}_{p n}
 -\epsilon\Omega^{2} \sin( \Omega t) \sum_{p} Q^{(k)}_{p,TE}(t)
g^{\ell}_{p n} \nonumber
\\ &+ \epsilon^{2}\Omega^{2} \cos^{2}(\Omega t)\sum_{p N} Q^{(k)}_{p,TE}(t) g^{\ell}_{p N}
 g^{\ell}_{n N}
 +\mathcal{O}(\epsilon^{2})
\label{ecqteapend}
\end{eqnarray}

It is well known that a naive perturbative solution of these
equations in powers of $\epsilon$ breaks down after a short amount
of time, of order $(\epsilon\Omega)^{-1}$. This happens for those
particular values of the external frequency $\Omega$ such that
there is a resonant coupling with the eigenfrequencies of the
static cavity. In this situation, to find a solution valid for
longer times (of order $(\epsilon^{-2} \Omega^{-1}) $) we use the
MSA technique \cite{Bender,crocce}. We introduce a second time
scale $\tau = \epsilon t$ and expand $Q^{(k)}_{n,TE}(t)$ as
follows
\begin{equation}
Q^{(k)}_{n,TE} =Q^{(k,0)}_{n,TE}(t,\tau) + \epsilon
Q^{(k,1)}_{n,TE}(t,\tau) +\mathcal{O}(\epsilon^{2}) \label{ap2}
\end{equation}
The initial conditions read
\begin{equation}
\ Q^{(k,0)}_{n,TE}(t=0)=\frac{\delta_{k n}}{\sqrt{2\omega_{\ell
k}}}\quad\quad \quad\quad
\dot{Q}^{(k,0)}_{n,TE}(t=0)=-\imath\sqrt{\frac{\omega_{\ell
k}}{2}}\delta_{k n} \label{ciD0}
\end{equation}

Replacing Eq.(\ref{ap2}) into Eq.(\ref{ecqteapend}), to zeroth
order in $\epsilon$ we get the equation of an harmonic oscillator.
The solution is
\begin{equation}
Q^{(k,0)}_{n,TE}(t,\tau)= A^{(k)}_{\ell n, TE}(\tau)e^{\imath
\omega_{\ell n}t} + B^{(k)}_{\ell n, TE}(\tau) e^{- \imath
\omega_{\ell n}t} \label{sol.QTE.0}
\end{equation}
and using the initial conditions it follows that
\begin{equation}
A^{(k)}_{\ell n, TE}(\tau=0)=0 \quad\quad \quad\quad B^{(k)}_{\ell
n, TE}(\tau=0)= \frac{1}{\sqrt{2\omega_{\ell k}}}\delta_{k n}
\label{ciD.AyB}
\end{equation}

To first order in $\epsilon$ we obtain
\begin{eqnarray}
&\partial_{tt}^{2}Q^{(k,1)}_{n,TE}(t,\tau)+ [\omega_{\ell
n}]^{2}Q^{(k,1)}_{n,TE}(t,\tau) = -2\partial_{\tau
t}Q^{(k,0)}_{n,TE}(t,\tau)+ 2\sin( \Omega t)[\omega_{\ell n}]^{2}
Q^{(k,0)}_{n,TE}(t,\tau)\nonumber
\\ &+
2\Omega \cos(\Omega t) \sum_{p}
\partial_{t}Q^{(k,0)}_{p,TE}(t,\tau) g^{\ell}_{p n} - \Omega^{2}
\sin( \Omega t) \sum_{p} Q^{(k,0)}_{p,TE}(t,\tau) g^{\ell}_{p n}
\label{QTE.primerorden}
\end{eqnarray}

The functions $A^{(k)}_{\ell n, TE}(\tau)$ and $B^{(k)}_{\ell n,
TE}(\tau)$ are obtained by imposing that no secular terms appear
in the equation for $Q^{(k,1)}_{n,TE}(t,\tau)$, that is, any term
with a time dependence of the form $e^{\pm i \omega_{\ell k} t}$
in the right-hand side of Eq.(\ref{QTE.primerorden}) must vanish.
We get
\begin{eqnarray}
&\partial_{\tau}A^{(k)}_{\ell n, TE} =-\frac{\omega_{\ell n}}{2}
B^{(k)}_{\ell n, TE}(\tau) \delta(\Omega-2 \omega_{\ell n}) +
\sum_{p\neq n} (\frac{\Omega}{2}-\omega_{\ell p})B^{(k)}_{\ell p,
TE}(\tau) g^{\ell}_{p n}\frac{\Omega}{2 \omega_{\ell n}}\nonumber
\\&\delta(\Omega-\omega_{\ell n}-\omega_{\ell p})
+\sum_{p\neq n} [(\omega_{\ell
p}+\frac{\Omega}{2})\delta(\Omega+\omega_{\ell p}-\omega_{\ell
n})+\nonumber
\\&
(\omega_{\ell p}-\frac{\Omega}{2})\delta(\Omega + \omega_{\ell
n}-\omega_{\ell p})] A^{(k)}_{\ell p, TE}(\tau) g^{\ell}_{p
n}\frac{\Omega}{2\omega_{\ell n}}
 \label{ecA}
\end{eqnarray}
\begin{eqnarray}
&\partial_{\tau} B^{(k)}_{\ell n, TE}=-\frac{\omega_{\ell n}}{2}
A^{(k)}_{\ell n, TE} \delta(\Omega-2\omega_{\ell n})+\sum_{p\neq
n} (\frac{\Omega}{2}-\omega_{\ell p})A^{(k)}_{\ell p, TE}
g^{\ell}_{p n}\frac{\Omega}{2\omega_{\ell n}}\nonumber
\\&\delta(\Omega-\omega_{\ell n}-\omega_{\ell p})+\sum_{p\neq
n} [(\omega_{\ell p}+\frac{\Omega}{2})\delta(\Omega+\omega_{\ell
p}-\omega_{\ell n})+ \nonumber
\\&
(\omega_{\ell p}-\frac{\Omega}{2})\delta(\Omega+\omega_{\ell
n}-\omega_{\ell p})]B^{(k)}_{\ell p, TE} g^{\ell}_{p
n}\frac{\Omega}{2 \omega_{\ell n}}
 \label{ecB}
\end{eqnarray}
The equations above lead to resonant behavior if
\begin{equation}
\Omega=2\omega_{\ell n} \quad\quad \textrm{or}\quad\quad
\Omega=\omega_{\ell n} +\omega_{\ell p} \label{res.par}
\end{equation}
These are the resonant conditions.  The modes $(\ell,n)$ and
$(\ell,q)$ are coupled if any of the following conditions is
satisfied
\begin{equation}
\Omega =\omega_{\ell n} -\omega_{\ell q} \quad\quad \Omega
=-\omega_{\ell n} +\omega_{\ell q} \label{acopl}
\end{equation}
These are the intermode coupling conditions.

The eigenfrequencies are determined by the zeros of the spherical
Bessel functions ($\omega_{\ell n}= \frac{\jmath_{\ell
n}}{a_{0}}$). For $\ell=0$ the spectrum is equidistant and
therefore the intermode coupling conditions are satisfied.
However, as for the electromagnetic field there are no modes with
$\ell=0$, in what follows we will consider only the case $\ell\neq
0$, in which the spectrum is not equidistant \cite{foot2}, and one
can check that the intermode coupling conditions are not
satisfied.

For the particular case $\Omega=2\omega_{LN}$,
Eqs.(\ref{ecA},\ref{ecB}) become
\begin{equation}
\partial_{\tau} A^{(k)}_{\ell n, TE}=-\frac{\omega_{\ell n}}{2} B^{(k)}_{\ell n, TE}
\delta_{\ell L} \delta_{nN}
 \label{ecAL}
\end{equation}
\begin{equation}
\partial_{\tau} B^{(k)}_{\ell n, TE}=-\frac{\omega_{\ell n}}{2} A^{(k)}_{\ell n, TE}
\delta_{\ell L} \delta_{nN}
 \label{ecBL}
\end{equation}
Using the initial conditions~(\ref{ciD.AyB}) we obtain, for
$\ell=L$ and $n=N$
\begin{equation}
A^{(k)}_{L N, TE}= \frac{-\delta_{k N}}{\sqrt{2\omega_{L N}}}
\sinh(\frac{\omega_{L N}}{2}\tau)
 \label{sol.A.L.d}
\end{equation}
\begin{equation}
B^{(k)}_{L N, TE}= \frac{\delta_{k N}}{\sqrt{2\omega_{L N}}}
\cosh(\frac{\omega_{L N}}{2}\tau) \label{sol.B.L.d}
\end{equation}
and for $\ell\neq L $
\begin{equation}
A^{(k)}_{\ell n, TE}= 0 \quad\quad\quad\quad B^{(k)}_{\ell n, TE}=
\frac{\delta_{k n}}{\sqrt{2 \omega_{\ell k}}}\label{sol.AyB.d}
\end{equation}

If the motion of the shell ends at $t=t_f$, the number of created
particles is given by
\begin{equation}
<\mathcal{N}_{N L m}>=<0_{in}|a^{out \dag} _{N L m}a^{out} _{N L
m}|0_{in}>= \sinh^{2}(\frac{\omega_{L N}}{2}\epsilon t_{f})
\end{equation}
\begin{equation}
<\mathcal{N}_{N L}>=\sum_{m} <0_{in}|a^{out \dag} _{N L m}a^{out}
_{N L m}|0_{in}>= (2L+1)\sinh^{2}(\frac{\omega_{L N}}{2}\epsilon
t_{f})
\end{equation}

In this Appendix we only considered the case of the scalar field
$\phi^{TE}$, associated to the TE modes.  The Eqs.(\ref{ec.Q.TM})
for the scalar field $\phi^{TM}$   can also be solved using MSA,
following the procedure described here.

%%%%%%%%%%%%%%%%%%%%%%%%%%%%%%%%%%%%%%%%%%%%%%%%%%%%%%%%%%%%%%%%%%%%%%%%%


\begin{thebibliography}{99}
\bibitem{foot1} There is an alternative way of deriving the
appropiate change of variables. From the definitions of $\eta$ and
$\xi$, using the chain rule it is easy to show that
$\partial_{\xi}=\partial_r -\partial_r g(r,t)\partial_t +
\mathcal{O}(\epsilon^{2})$. The conditions in Eq. (\ref{condg})
imply that, at $r=a(t)$ and up to order $\mathcal{O}(\epsilon^{2})$,
$\partial_{\xi}=\partial_r -\dot a\partial_t$ and $\eta = t$.
Therefore, the usual boundary condition in the new coordinates
Eq.(\ref{esc.Neumann.eta.xi}) is equivalent to the generalized one
in the old coordinates.
\bibitem{foot2} Due to the properties of the zeroes of the Bessel functions,
for $l\neq 0$ and large values of $n$ the spectrum becomes almost
equidistant. However, as a small detuning breaks the parametric
resonance and the mode coupling \cite{crocce}, in order to have
intermode coupling with $l\neq 0$ it is necessary to consider very
large values of $n$,  and therefore uninterestingly large external
frecuencies.
\end{thebibliography}
\end{document}